\documentstyle[12pt]{article}
\input epsf
\parskip=\medskipamount
\def\caH{{\cal H}}
\def\al{\alpha}
\def\be{\beta}

\def\la{\lambda}  \def\La{\Lambda}
   
\def\om{\omega}   
   
\def\IC{\relax{\rm l\kern-.50 em C}}
\def\IE{\relax{\rm l\kern-.12 em E}}
\def\IK{\relax{\rm l\kern-.18 em K}}
\def\IL{\relax{\rm I\kern-.18 em L}}
\def\IN{\relax{\rm I\kern-.18 em N}}
\def\IR{\relax{\rm I\kern-.18 em R}}

\font\tenfrak=eufm10  \font\sevenfrak=eufm7  \font\fivefrak=eufm5
\newfam\frakfam
\textfont\frakfam=\tenfrak
\scriptfont\frakfam=\sevenfrak
\scriptscriptfont\frakfam=\fivefrak

\newtheorem{proposicion}{Proposition}

\def\wt{\widetilde}
\def\frac#1#2{{#1\over #2}}
\def\fracpd#1#2{\frac{\partial #1}{\partial #2}}

\def\ket#1{|#1\rangle}
\begin{document}

\title{A Quantum Exactly Solvable Nonlinear Oscillator  \\
with quasi-Harmonic Behaviour }

\author{
Jos\'e F. Cari\~nena$\dagger\,^{a)}$,
Manuel F. Ra\~nada$\dagger\,^{b)}$,
Mariano Santander$\ddagger\,^{c)}$ \\
$\dagger$
    {\sl Departamento de F\'{\i}sica Te\'orica, Facultad de Ciencias} \\
    {\sl Universidad de Zaragoza, 50009 Zaragoza, Spain}  \\
$\ddagger$
    {\sl Departamento de F\'{\i}sica Te\'orica, Facultad de Ciencias} \\
    {\sl Universidad de Valladolid,  47011 Valladolid, Spain}
}
\maketitle
\date{}

\begin{abstract}
The quantum version of a non-linear oscillator, previouly analyzed
at the classical level, is studied. This is a problem of
quantization of a system with position-dependent mass of the form
$m={(1+\lambda x^2)}^{-1}$ and with a $\la$-dependent
nonpolynomial rational potential. This $\la$-dependent system can
be considered as a deformation of the harmonic oscillator in the
sense that for $\la\to 0$ all the characteristics of the linear
oscillator are recovered. Firstly, the $\la$-dependent
Schr\"odinger equation is exactly solved as a Sturm-Liouville
problem and the $\la$-dependent eigenenergies and eigenfunctions
are obtained for both $\la>0$ and $\la<0$. The $\la$-dependent
wave functions appear as related with a family of orthogonal
polynomials that can be considered as $\la$-deformations of the
standard Hermite polynomials. In the second part, the
$\la$-dependent Schr\"odinger equation is solved by using the
Schr\"odinger factorization method, the theory of intertwined
Hamiltonians and the property of shape invariance as an approach.
Finally, the new family of orthogonal polynomials is studied.
We prove the existence of a $\la$-dependent Rodrigues formula,
a generating function and $\la$-dependent recursion relations
between polynomials of different orders.
\end{abstract}

\begin{quote}
Position-dependent mass. Sch\-r\"o\-dinger equation. Hermite polynomials.
Intertwined Hamiltonians. Shape-inva\-riant potentials.


Schr\"odinger, Dirac,
\end{quote}
{\vfill}

\footnoterule
{\noindent\small
$^{a)}${\it E-mail address:} {jfc@unizar.es}  \\
$^{b)}${\it E-mail address:} {mfran@unizar.es} \\
$^{c)}${\it E-mail address:} {santander@fta.uva.es}
\newpage

\section{Introduction }

  The nonlinear differential equation
\begin{equation}
   (1 +\la x^2)\,\ddot{x} - (\la x)\,\dot{x}^2 + \al^2\,x  = 0
   \,,\quad\la>0\,,             \label{Eq1}
\end{equation}
was studied by Mathews and Lakshmanan in \cite{MaLa74} (see also
\cite{LaRa03}) as an example of a non-linear oscillator (notice
$\al^2$ was written just as $\al$ in the original paper); the most
remarkable property is the existence of solutions of the form
$$
  x = A \sin(\om\,t + \phi) \,,
$$
with the following additional restriction linking frequency and
amplitude
$$
  \om^2 = \frac{\al^2}{1 + \la\,A^2} \,.
$$
That is, the equation (\ref{Eq1}) represents a non-linear
oscillator with periodic solutions that were qualified as having a
``simple harmonic form". The authors also proved that (\ref{Eq1})
is obtainable from the Lagrangian
\begin{equation}
   L = \frac{1}{2}\,\Bigl(\frac{1}{1 + \la\,x^2}\Bigr)\,
   (\dot{x}^2 - \al^2\,x^2)       \label{LagClas}
\end{equation}
which they considered as the one-dimensional analogue of the
Lagrangian density
$$
  {\cal L} = \frac{1}{2}\,\Bigl(\frac{1}{1 + \la\,\phi^2} \Bigr)\,
  (\partial_{\mu}\phi\,\partial^{\mu}\phi - m^2\,\phi^2)    \,,
$$
appearing in some nonpolynomial models of quantum field theory.

   The nonlinear equation (\ref{Eq1}) is therefore an interesting
example of a system with nonlinear quasi-harmonic oscillations.
Recently, it has been proved \cite{CaRaSS04} that this particular
nonlinear system can be generalized to the two-dimensional case,
and even to the $n$-dimensional case and that these higher
dimensional systems are superintegrable with $2n-1$ quadratic
constants of motion. Moreover, we point out that a geometric
interpretation of the higher-dimensional systems was proposed
in relation with the dynamics on spaces of constant curvature.
It was also proved the existence of a related
$\la$-dependent isotonic oscillator and that the two-dimensional
oscillator, previously studied in \cite{CaRaSS04}, admits a
superintegrable modification that corresponds to the
$\la$-dependent version of the Smorodinski-Winternitz sytem
\cite{CaRaS05,CaRaS06}. In fact this means that the deformation
introduced by the parameter $\la$ modifies the Hamilton-Jacobi
equation but preserves the existence of a multiple separability.

   On the other hand, Biswas {\sl et al} studied in 1973 \cite{BiDaS73} the
ground state as well  as some excited energy levels of the
generalized anharmonic oscillator defined by the Hamiltonian
$H_m=-\,d^2/dx^2+x^2+\la x^{2m}$, $m=2,3,\dots$, and then they
proposed the use of similar techniques for the analysis of the
Schr\"odinger equation involving the potential $\la(x^2/(1 + g
x^2))$. Since then, this nonpolynomial potential has been
extensively studied by many authors \cite{MaLa75}-\cite{Is02} from
different viewpoints and by making use of different approaches. In
many cases the term $x^2/(1 + g x^2)$ was introduced as a
perturbation of an initial harmonic oscillator; that is, the
potential to be solved was not $x^2/(1 + g x^2)$ by itself but
$x^2+\la x^2/(1+gx^2)$ (exact analytical solutions have only been
found for certain very particular values of the parameters $\la$
and $g$, see e.g., Refs. \cite{BeBe80}--\cite{Zn99Jpa}). It is
important to note that, in most of these cases, the derivative
part of the Schr\"odinger equation was the standard one, that is,
the equation arising from a Hamiltonian $H$ with a quadratic term
of the form $(1/2)p^2$ and leading to a derivative term of the
form $-\,d^2/dx^2$, that is
$$
  \Bigl[-\,\frac{d^2}{dx^2} + x^2+\la \frac{x^2}{(1+gx^2)}\Bigr]\,\Psi_n
  = E_n\,\Psi_n
$$
or the corresponding two or three-dimensional versions involving
the Laplace operator in $\IE^2$ or $\IE^3$.

  The important point is that, in the Lagrangian (\ref{LagClas}), the
parameter $\la$ is present not only in potential $x^2/(1 + \la\,
x^2)$ but also in the kinetic term. So, this nonlinear oscillator
must be considered as a particular case of a system with a
position-dependent effective mass \cite{Le95}. In fact, it is
known the existence of physical systems for which the mass of a
nonrelativistic quantum particle may vary with its position (they
correspond, for example, to the motion in an external field in a
crystal, electronic properties of semiconductors, liquid crystals,
etc.) and in recent years special interest has been drawn to
generalize the standard methods of solving the Schr\"odinger
equation to this new class of systems
\cite{DeChH98}-\cite{JiYiJ05}. For example, some techniques have
been proposed to convert the Schr\"odinger equation with a
position-dependent mass into a Schr\"odinger equation with a
constant mass (see \cite{JiYiJ05} and references therein).
Nevertheless there is an important problem at the starting level
of quantization (that is, transition from the classical system to
the quantum one) since if the mass $m$ becomes a spatial function,
$m=m(x)$, then the quantum version of the mass no longer commutes
with the momentum. Therefore, different forms of presenting the
kinetic term in the Hamiltonian $H$,  as for example
$$
  T = \frac{1}{4}\,\Bigl[\frac{1}{m(x)}\,p^2 + p^2\,\frac{1}{m(x)}\Bigr]\,,\quad
  T = 
\frac{1}{2}\,\Bigl[\frac{1}{\sqrt{m(x)}}\,p^2\,\frac{1}{\sqrt{m(x)}}\Bigr]\,,\quad
  T = \frac{1}{2}\,\Bigl[p\,\frac{1}{m(x)}\,p \Bigr]\,,
$$
are equivalent at the classical level but they lead to different
nonequivalent Schr\"odinger equations.

   In fact, the quantum version of Mathews and Lakshmanan oscillator
was studied in \cite{MaLa75} by making use of the following Hamiltonian
$$
  H = \frac{1}{2} \Biggl[\frac{1}{2} \bigl\{ p^2\,,\,(1-gx^2) \bigr\}
  + \frac{k\,x^2}{(1-gx^2)}\Biggr]\,,
$$
where the notation $\{A,B\} = A B + B A$ is used. More recently
another different approach has been discussed in Ref. \cite{CaRaS04}.
It was proposed a rule for the transition from the
classical system to the quantum one that was obtained from an
analysis of the geometric properties of the function representing
the classical kinetic energy. One of the advantages of this
procedure was that the quantum system so obtained can be studied
by using a factorization method \cite{Sch40}-\cite{CaRa00Rmp} and
that it is even endowed with the property of shape invariance
\cite{Ge83}-\cite{SaZa05}.

   The main objective of this article is to continue with the analysis
of the quantum version of this particular nonlinear oscillator and
to present a detailed study not only of the central question but also of
some questions related with it.
The main points to be discussed in this paper can be summarized
in the following four points:
\begin{itemize}
\item{} Analysis of the transition from the classical $\la$-dependent
system to the quantum one.

This is a problem of quantization of a system with position-dependent
mass and in this particular case the quantization procedure to be
analyzed is related with the existence, at the classical level,
of a Killing vector and also of a $\la$-dependent invariant
measure.

\item{} Analysis and exact resolution of the $\la$-dependent Schr\"odinger
equation as a Sturm-Liouville problem.

In fact, not just one but two different Sturm-Liouville problems
are obtained: one for $\la$ positive and other for $\la$ negative.
This also means a discussion of the properties of the energy levels and of
wave functions depending of the sign of $\la$.

\item{} Existence of a family of $\la$-dependent orthogonal polynomials.

The $\la$-dependent system we study can be considered as a deformation of
the harmonic oscillator in the sense that for $\la\to 0$ all the
characteristics of the linear oscillator must be recovered.
The $\la$-dependent wave functions will appear as
related with a family of polynomials that can be considered as
$\la$-deformations of the standard Hermite polynomials. This means
that the resolution of the quantum problem leads in a natural way
to the study of a new family of orthogonal polynomials.

\item{} Resolution of the $\la$-dependent Schr\"odinger equation
by using the Schr\"odinger factorization method as an approach.

This means a discussion of this $\la$-dependent quantum problem
as a shape-invariant problem.
Many properties of the $\la$-dependent orthogonal polynomials,
as the generating function and some recursion relations, are
obtained as a byproduct of this factorization approach.

\end{itemize}

    In more detail, the plan of the article is as follows:
In Sec. II we first discuss the quantization of the classical
system and then we solve the $\la$-dependent Schr\"odinger equation;
we analyze the characteristics of the solutions and we prove that
they depend on a family of polynomials related with the Hermite polynomials;
in the last subsection we prove the orthogonality of these new polynomials.
Sec. III is devoted to review the main characteristics of the
Schr\"odinger factorization formalism and in Sec. IV we apply this
method to the $\la$-dependent nonlinear oscillator and we obtain, using the
operators $A$ and $A^+$, the energies $E_n$ and the wave functions
$\Psi_n$.
In Sec. V we study the main properties of the $\la$-dependent
``deformed Hermite" polynomials; in particular we prove the
existence of a $\la$-dependent Rodrigues formula, of a generating
function and of $\la$-dependent recursion relations between polynomials
of different orders.
Finally, in Sec. VI we discuss the results and make some final comments.

\section{$\la$-dependent Schr\"odinger equation }

\subsection{Quantization and Schr\"odinger equation}

In the general case of a system with position-dependent effective
mass the transition from the classical system to the quantum one
is a difficult problem because of the ambiguities in the order of
the factors. Nevertheless, in the particular case of the
Lagrangian (\ref{LagClas}), it was proved in Ref. \cite{CaRaS04}
that the structure of the kinetic term suggests us a very clear and
direct procedure for the quantization.

   Let us begin by considering the one-dimensional free-particle
motion characterized by the $\la$-dependent kinetic term $T(\la)$
as a Lagrangian
\begin{equation}
  L(x,v_x;\la) = T_\la(x,v_x)
  = \frac{1}{2}\,\Bigl(\frac{v_x^2}{1 + \la\,x^2}\Bigr)\,,
\end{equation}
and the following nonlinear equation
$$
  (1 +\la x^2)\,\ddot{x} - (\la x)\,\dot{x}^2 = 0 \,,
$$
where we point out that we admit $\la$ can take both positive and
negative values; of course it is clear that for $\la<0$,
$\la=-\,|\la|$, the function (and the associated dynamics) will
have a singularity at $1 -\,|\la|\,x^2=0$ and   we
shall restrict the study of the dynamics to the interior of the
interval $x^2<1/|\la|$ where the kinetic energy function $T_\la$ is
positive definite.

The function $T_\la$ is invariant
under the action of the vector field $X(\la)$ given by
$$
  X_\la(x) = \sqrt{\,1+\la\,x^2\,}\,\,\fracpd{}{x}  \,,
$$
in the sense that we have
$$
  X^t_\la\Bigl(T_\la\Bigr)=0  \,,
$$
where $X^t_\la$ denotes the natural lift to the velocity phase space
$\IR{\times}\IR$ (the tangent bundle in differential geometric terms) of the
vector field $X_\la$,
$$
  X^t_\la = \sqrt{\,1+\la\,x^2}\,\,\fracpd{}{x}
  + \Bigl(\frac{\la\,x v_x}{\sqrt{1+\la\,x^2}}\Bigr)\fracpd{}{v_x} \,.
$$
In differential geometric terms this property means that the
vector field $X_\la$ is a Killing vector of the one-dimensional
metric
$$
  ds_\la^{2} = \Bigl(\frac{1}{1 + \la\,x^2}\Bigr)\,dx^2 \,.
$$
Moving to the quantum setting, we see that this property suggests
the idea of working with functions and linear operators defined on
the space obtained by considering the real line $\IR$ endowed with
the measure $d\mu_\la$ given by
\begin{equation}
   d\mu_\la = \Bigl(\frac{1}{\sqrt{1+\la\,x^2}}\Bigr)\,dx \,,\label{dmu}
\end{equation}
which is (up to a factor) the only measure invariant under
$X_\lambda$ \cite{CaRaS04}. This means that the operator $P_x$
representing the linear momentum must be self-adjoint not in the
standard space $L^2(\IR)$ but in the space $L^2(\IR,d\mu_\la)$.

Let us start our study with the classical Hamiltonian of the
$\la$-dependent oscillator \cite{CaRaSS04}
\begin{equation}
   H = \bigl(\frac{1}{2 m}\bigr)\,P_x^2
   + \bigl(\frac{1}{2}\bigr)\,g\, \Bigl(\frac{x^2}{1 + \la\,x^2}\Bigr)  \,,
   \quad P_x = \sqrt{1 + \la\,x^2}\,p_x  \,,
   \quad g = m\al^2  \,.   \label{HClas}
\end{equation}
Figures I and II show the form of the potential $V_{\la}(x)$ for
several values of $\la$ ($\la<0$ in Figure I and $\la>0$ in Figure
II).

As it has been pointed out in \cite{CaRaS04}, the generator of the
infinitesimal 'translation' symmetry $\sqrt{1 + \la\,x^2}
\,{d}/{dx}$ is skew-selfadjoint in the space $L^2(\IR,d\mu_\la)$
and therefore the transition from the classical system to the
quantum one is given by defining the operator
$$
  P_x =\ -\,i\,\hbar\,\sqrt{1 + \la\,x^2}\,\frac{d}{dx}  \,,
$$
so that
$$
  (1 + \la\,x^2)\,p_x^2 \ \to\ -\,\hbar^2\,
  \Bigl(\sqrt{1 + \la\,x^2}\,\frac{d}{dx}\Bigr)
  \Bigl(\sqrt{1 + \la\,x^2}\,\frac{d}{dx}\Bigr) \,,
$$
in such a way that the quantum version of the Hamiltonian
(\ref{HClas}) becomes
$$
  \widehat{H} = - \frac{\hbar^2}{2 m}\,(1 + \la\,x^2)\,\frac{d^2}{dx^2}
  - \bigl(\frac{\hbar^2}{2 m}\bigr)\,\la\,x\,\frac{d}{dx}
  + \bigl(\frac{1}{2}\bigr)\,g \Bigl(\frac{x^2}{1 + \la\,x^2}\Bigr) \,.
$$

Let us consider the following quantum Hamiltonian
\begin{equation}
  \widehat{H} = -\,\frac{\hbar^2}{2 m}\,(1 + \la\,x^2)\,\frac{d^2}{dx^2}
  - \bigl(\frac{\hbar^2}{2 m}\bigr)\,\la\,x\,\frac{d}{dx}
  + \bigl(\frac{1}{2}\bigr)\,m\al\,(\al +
  \frac{\hbar}{m}\,\la)\Bigl(\frac{x^2}{1 + \la\,x^2}\Bigr) \,,
\label{Hq(xla)}
\end{equation}
where we have slightly modified the value of the parameter $g$
that now is given by $g = m\al^2 + \la\,\hbar\al$; this is done in
order that the notation used in this approach coincides with the
one to be presented in the next section III. It is also convenient
to simplify this function $\widehat{H}$ by introducing
adimensional variables $(y,\La)$ defined by
$$
  x = \Bigl(\sqrt{\frac{\hbar}{m\al}}\,\Bigr)\,y \,,\quad
  \la = \Bigl(\frac{m\,\al}{\hbar}\Bigl)\,\La  \,,
$$
in such a way that the following equality is satisfied
$$
  1 + \la\,x^2 = 1 + \La\,y^2  \,.
$$
The Hamiltonian $\widehat{H}$ takes then the following form
\begin{equation}
\widehat{H} = \Bigl[\,
   - \frac{1}{2}\,(1 + \La\,y^2)\,\frac{d^2}{dy^2}
   - \bigl(\frac{1}{2}\bigr)\,\La\,y\,\frac{d}{dy}
   + \bigl(\frac{1}{2}\bigr)\,(1 + \La)\Bigl(\frac{y^2}{1 + \La\,y^2}\Bigr)
   \,\Bigr] \bigl(\hbar\,\al\bigr)\,,
\end{equation}
and then the  Schr\"odinger equation
$$ \widehat{H}\,\Psi = E\,\Psi \,,\qquad  E = e\,(\hbar\,\al) \,,
$$
reduces to the following adimensional form
$$
\Bigl[\,
   - \frac{1}{2}\,(1 + \La\,y^2)\,\frac{d^2}{dy^2}
   - \bigl(\frac{1}{2}\bigr)\,\La\,y\,\frac{d}{dy}
   + \bigl(\frac{1}{2}\bigr)\,(1 + \La)\Bigl(\frac{y^2}{1 + \La\,y^2}\Bigr)
   \,\Bigr]\,\Psi  = e\,\Psi  \,,
$$
which, after a small simplification, can be finally rewritten as follows
\begin{equation}
  (1 + \La\,y^2) \frac{d^2}{dy^2}\,\Psi + \La y\,\frac{d}{dy}\,\Psi
  - (1 + \La)\Bigl(\frac{y^2}{1 + \La\,y^2}\Bigr)\,\Psi + (2\,e)\,\Psi
  = 0  \,.
\label{EqPsi(yLa)}
\end{equation}

  It is known that in the $\La=0$ case the asymptotic behaviour at the
infinity suggest a factorization for the wave function.
The idea is that a similar procedure can be applied to this 
$\La$-dependent equation.
Let us first denote by $\Psi_{\infty}$ the following function
$\Psi_{\infty}=(1 + \La\,y^2)^{-\,1/(2\La)}$
that vanish in the limit $y^2\to\infty$ (in the case $\La>0$)
or in the limit $y^2\to -1/\La$ (in the case $\La<0$). Then we obtain
$$
  \Biggl[(1 + \La\,y^2) \frac{d^2}{dy^2} + \La y\,\frac{d}{dy}
  - (1 + \La)\Bigl(\frac{y^2}{1 + \La\,y^2}\Bigr)  \Biggr]\Psi_{\infty}
  = -\,\Psi_{\infty}  \,.
$$
Thus, $\Psi_{\infty}$ is the exact solution in the very particular case $e=1/2$
and represents the asymptotic behaviour of the solution in the general case.
Consequently, this property suggest the following factorization
\begin{equation}
   \Psi(y,\La) = h(y,\La)\,(1 + \La\,y^2)^{-\,1/(2\La)} \label{Psi(yLa)}\,,
\end{equation}
and then the new function $h(y,\La)$ must satisfy the differential
equation
\begin{equation}
   (1 + \La\,y^2) h'' + (\La - 2) y h' + (2 e -1) h = 0 \,,\qquad
h=h(y,\La)\,.\label{Eqh(yLa)}
\end{equation}
Let us first remark that, as $\lim_{\La\to 0}(1+\La\, y^2)=1$ and
$\lim_{\La\to 0 }[1/(2\,\La)]=\infty$, we have
$$
  \lim_{\La\to 0}(1 + \La\,y^2)^{1/(2\La)}
  = \exp\left[\lim_{\La\to 0} \frac{\log(1+\La y^2)}{2\,\La}\right]
  = \exp\left[\lim_{\La\to 0}\frac{y^2/(1+\La\,y^2)}{2}\right] = e^{y^2/2} \,.
$$
Consequently,
$$
  \lim{}_{\La{\to}0}\,\Psi(y,\La) = h(y,0)\,e^{-\,(1/2)\,y^2}\,.
$$

Furthermore, if we put $\La=0$ in the equation (\ref{EqPsi(yLa)})
we see that $\bar h(y)=h(y,0)$ satisfies the Hermite equation
$$
  \La\,\to\,0 {\qquad} \bar{h}''- 2\,y\,\bar{h}' + (2 e -1)\,\bar{h}
  = 0
$$
In other words, for  $\La=0$, the  function $h(y,0)$ is a solution
of  Hermite equation. Recall that when $2 e -1$ is an even integer
number,  i.e. $2 e -1= 2p$, such differential equation admits
polynomial solutions, called (with an appropriately chosen overall
factor) Hermite polynomials.

Nevertheless, it is also interesting to note that the dependence
of the parameter $\La$ makes of (\ref{Eqh(yLa)}) an equation with
certain similarities with the Legendre's equation.

  It is clear that the three coefficient functions of the linear
differential equation
$$
  (1 + \La\,y^2) h'' + (\La - 2) y h' + (2 e - 1) h = 0 \,,\qquad
  h= h(y,\La)\,,
$$
are analytic at the origin (and the director coefficient does not
vanish in a neighbourhood of $y=0$). The origin is therefore an
ordinary point and we expect an analytic solution $h$ with a power
series expansion convergent in an interval $(-R,R)$ for some non-zero
value of the radius of convergence $R$,
$$
  h(y,\La) = \sum_{n=0}^\infty\,a_n(\La)\,y^n
   = a_0(\La) + a_1(\La)\,y + a_2(\La)\,y^2 + \dots
$$
In this way the equation becomes
\begin{equation}
  \sum_{n=0}^\infty\,\left[ (n+2)(n+1)\,a_{n+2}+ \La\,a_n\,y^n
   + (\La - 2)\,n\,a_n + (2 e -1)\,a_n \right]\,y^n = 0  \,,
\end{equation}
and therefore it determines to the following $\La$-dependent
recursion relation
$$
  a_{n+2} = (-1)\,\frac{a_n}{(n+2)(n+1)}\,
  \Bigl[\,n\,(\La\,n -\,2) + (2 e -1)\,\Bigr]\,,\qquad  n=0,1,2,\ldots
$$
Note that this relation shows that, as in the $\La=0$ case, even
power coefficients are related among themselves and the same is true for
odd power coefficients. In both cases, having in mind that
$$
\lim{}_{n\to\infty}\,\biggl|\frac{a_{n+2}x^{n+2}}{a_nx^n}\biggr|
   = \lim\nolimits_{n\to \infty}\,\biggl|\,\frac{\,n\,(\La\,n -\,2)
   + (2 e -1)\,}{(n+2)(n+1)}\,\biggr|\,\bigl|\,x^2\bigr| =
|\,\La\,|\,\bigl|\,x^2\bigr|
$$
so that the radius of convergence $R$ is therefore given by
$$
  R = \frac{1}{\sqrt{\,|\,\La\,|\,}} \,.
$$
Hence, when we consider the limit $\La\to0$, we
recover the radius $R=\infty$ of the Hermite's equation.

The general solution $h$ is given by the linear combination  $h =
a_0 h_1 + a_1 h_2$ where  $h_1(y)$ and  $h_2(y)$ are the solutions
determined by $h_1(0)=1,\ h_1'(0)=0$ and $h_2(0)=0,\ h_2'(0)=1$,
respectively.

The condition for the differential equation to admit a polynomial
solution is the existence of an integer number $p$ such that
$$
  2 e -1 = 2\,p - \La\,p^2 \,,
$$
because then
$$
  a_p {\ne}0\,,\quad a_{p+2} = 0 \,,
$$
and the solution is a polynomial of order $p$.
This relation establishes some possible values for the spectrum of the
Hamiltonian $\widehat H$ and therefore of $H$ (this question will be
discussed in the next subsection).

The polynomial solutions are given by
\begin{itemize}
\item{} Even index (even power polynomials)
\begin{eqnarray}
  {\cal H}_{2p} &=& k_{2p} \sum_{r=0}^{r=p} \,a_{2r}\,y^{2r} \cr
  a_{2r} &=& (-1)^r\,\frac{a_0}{2r\,!}\,p'\,(p'-2)(p'-4)\dots(p'-2(r-1)) \cr
  &&{\hskip50pt}\bigl[\,2 - \La\,p'\,\bigr]\bigl[\,2 - \La (p'+2)\,\bigr]
  \bigl[\,2 - \La(p'+4)\,\bigr]\dots\bigl[\,2 - \La\,(p'+2(r-1))\,\bigr]
\nonumber \end{eqnarray}
where we have introduced the notation $p'=2p$. More specifically,
the expressions of the first solution $h_1$, in the particular
cases of $p'=0,2,4,6$, are given by:
\begin{eqnarray}
  {\cal H}_0 &=& k_0  \cr
  {\cal H}_2 &=& k_2\,[ 1 - 2(1 - \La)y^2\,] \cr
  {\cal H}_4 &=& k_4\,[ 1 - 4(1 - 2\La)y^2 +
   (\frac{4}{3})(1 - 2\La)(1 - 3\La)y^4 ] \cr
  {\cal H}_6 &=&  k_6\,[ 1 - 6(1 - 3\La)y^2 + 4(1 - 3\La)(1 - 4\La)y^4
   + (\frac{8}{15})(1 - 3\La)(1 - 4\La)(1 - 5\La)y^6 ] \nonumber
\end{eqnarray}
where $k_0$, $k_1$, $k_2$, $\dots$, are overall multiplicative constants.
\item{} Odd index  (odd power polynomials)
\begin{eqnarray}
   {\cal H}_{2p+1} &=& k_{2p+1} \sum_{r=0}^{r=p} \,a_{2r+1}\,y^{2r+1} \cr
   a_{2r+1} &=& 
(-1)^{r+1}\,\frac{a_1}{2r+1\,!}\,(p'-1)(p'-3)\dots(p'-(2r-1)) \cr
   &&{\hskip80pt}\bigl[\,2 - \La\,(p'+1)\,\bigr]\bigl[\,2 - \La (p'+3)\,\bigr]
   \dots\bigl[\,2 - \La\,(p'+(2r-1))\,\bigr]\nonumber
\end{eqnarray}
where we have introduced the notation $p'=2p+1$. More specifically,
the expressions of the second solution $h_2$ for
$p'=1,3,5$, are given by:
\begin{eqnarray}
  {\cal H}_1 &=& k_1\,y  \cr
  {\cal H}_3 &=& k_3\,[ y - (\frac{2}{3})(1 - 2\La)y^3 ]    \cr
  {\cal H}_5 &=& k_5\,[ y - (\frac{4}{3})(1 - 3\La)y^3
   + (\frac{4}{15})(1 - 3\La)(1 - 4\La)y^5 ]   \nonumber
\end{eqnarray}
where $k_1$, $k_3$, $k_5$, $\dots$, are arbitrary multiplicative constants.
\end{itemize}
It is clear that these particular polynomials solutions ${\cal H}_m$,
$m=0,1,2,\dots,$ play, for this $\La$-dependent oscillator, a
similar role to the Hermite's polynomials for the standard
harmonic oscillator.

\subsection{$\la$-dependent Sturm-Liouville problem and orthogonality}

The $\La$-dependent differential equation
$$
  a_0 h'' + a_1 h' + a_2 h = 0 \,,
$$
$$
  a_0 = (1 + \La\,y^2) \,,\quad a_1 = (\La - 2)\,y
  \,,\quad a_2 = (2 e -1) \,,\quad
$$
is not self-adjoint since $a'_0 \ne a_1$ but it can be reduced to
self-adjoint form by making use of the following integrating
factor
$$
  \mu(y) = (\frac{1}{a_0})\,e^{{\int}(a_1/a_0)\,dy}
  = (1 + \La\,y^2)^{-(\La +2)/(2\,\La)}
$$
in such a way that we arrive to the following expression
$$
  \frac{d}{dy}\Bigl[\,p(y,\La)\,\frac{dh}{dy}\,\Bigr]
  + (2 e -1)\,r(y,\La)\,h = 0 \,,
$$
where the two functions $p=p(y,\La)$ and $r=r(y,\La)$ are given by
\begin{eqnarray}
   p(y,\La) &=& e^{{\int}(a_1/a_0)\,dy} = (1 + \La\,y^2)^{1/2 - 1/\La} 
\,,\cr&&\cr
   r(y,\La) &=& (\frac{a_2}{a_0})\,e^{{\int}(a_1/a_0)\,dx}
   = \frac{1}{1 + \La\,y^2} (1 + \La\,y^2)^{1/2 - 1/\La} \,.
\nonumber\end{eqnarray}
Thus, we have obtained
\begin{equation}
  \frac{d}{dy}\Bigl[\,\Bigl(\,\frac{\sqrt{1 + \La\,y^2}}
  {(1 + \La\,y^2)^{1/\La}}\,\Bigr) \,\frac{dh}{dy} \,\Bigr]
  + \frac{(2 e -1)\,h}{\sqrt{1 + \La\,y^2}\,(1 + \La\,y^2)^{1/\La}} = 0
\label{ESL(yLa)}
\end{equation}
that, together with appropriate conditions for the behaviour of
the solutions at the end points, constitute a  Sturm-Liouville
problem. It is to be pointed out that the boundary conditions are
in fact different according to the sign of $\La$; therefore we
arrive to, no just one, but two different Sturm-Liouville
problems:
\begin{itemize}
\item{} If $\La$ is negative the range of the  variable $y$ is
limited by the restriction $y^2<1/|\La|$. In this case the
problem,  defined in the bounded interval
$[-\,a_\La,a_\La]$ with $a_\La=1/\sqrt{|\La|}$, is singular because
the function $p(y,\La)$ vanishes in the two end points
$y_1=-\,a_\La$ and $y_2=a_\La$. The conditions to be imposed in
this case lead to prescribe that the solutions $h(y,\La)$ of the
problem must be bounded functions at the two end points,
$y_1=-\,a_\La$ and $y_2=a_\La$, of the interval.
It is clear that this leads to the above mentioned polynomial
solutions.

\item{} If $\La$ is positive  the variable $y$ is
defined in the whole real line $\IR$ and, therefore, the
Sturm-Liouville problem is singular. The solutions $h(y,\La)$ must
be well defined in all $\IR$,
and the boundary conditions prescribe that the behaviour of these
functions when $y\to\pm\,\infty$ must be such that their norms,
determined with respect to the weight function $r(y)$, be finite.
It is clear that in this case the solutions of the problem
are again the $\La$-dependent polynomials ${\cal H}_m$, $m=0,1,2,\dots$
\end{itemize}

\begin{proposicion}
The eigenfunctions of the problem (\ref{ESL(yLa)}) are orthogonal
with respect to the function $r= (1 + \La\,y^2)^{-(1/2 + 1/\La)}$.
\end{proposicion}
{\it Proof:} This statement is just a consequence of the
properties of the Sturm-Liouville problems.

Because of this the polynomial solutions ${\cal H}_m$,
$m=0,1,2,\dots$, of the equation (\ref{Eqh(yLa)}), satisfy
\begin{equation}
   \int_{-\,a_\La}^{a_\La} \frac{{\cal H}_m(y,\La)\,{\cal H}_n(y,\La)}
   {(1 + \La\,y^2)^{1/\La}\,\sqrt{1 + \La\,y^2}} \,\, dy = 0
   \,,\quad m\,\ne\,n \,,\quad \La<0\,,
\end{equation}
and
\begin{equation}
   \int_{-\infty}^{\infty} \frac{{\cal H}_m(y,\La)\,{\cal H}_n(y,\La)}
   {(1 + \La\,y^2)^{1/\La}\,\sqrt{1 + \La\,y^2}} \,\, dy = 0
   \,,\quad m\,\ne\,n \,,\quad \La>0\,.
\end{equation}

If we define the {\sl $\La$-dependent Hermite functions\/}
$\Psi_m$ by
$$
  \Psi_m(y,\La) = {\cal H}_m(y,\La)\,(1 + \La\,y^2)^{-\,1/(2\La)}
  \,,\quad  m=0,1,2,\dots
$$
then the above statement admits the following alternative form:
{\sl The $\La$-dependent Hermite functions $\Psi_m(y,\La)$ are
orthogonal with respect to the weight function  $\wt{r}=1/\sqrt{1
+ \La\,y^2}$}:
\begin{equation}
  \int_{-\,a_\La}^{a_\La}\Psi_m(y,\La)\,\Psi_n(y,\La)\,\wt{r}(y,\La) \,dy
  = \int_{-\,a_\La}^{a_\La}\Psi_m(y,\La)\,\Psi_n(y,\La)\frac{dy}
  {\sqrt{1 + \La\,y^2}} = 0
  \,,\quad m\,\ne\,n \,,\quad \La<0\,,
\end{equation}
and
\begin{equation}
  \int_{-\infty}^{\infty}\Psi_m(y,\La)\,\Psi_n(y,\La)\,\wt{r}(y,\La) \,dy
  = \int_{-\infty}^{\infty}\Psi_m(y,\La)\,\Psi_n(y,\La)\frac{dy}
  {\sqrt{1 + \La\,y^2}} = 0
  \,,\quad m\,\ne\,n \,,\quad \La>0\,.
\end{equation}

Note that this orthogonality with respect to the weight function
$\wt{r}$ coincides with the orthogonality with respect to the
measure $d\mu_\la$ discussed in the subsection 2.1.

Finally, we also remark that
$$
  \lim{}_{\La\to 0} \Psi_m(y,\La) = H_m(y)\,e^{-\,(1/2)\,y^2}
  \,,\quad m=0,1,2,\dots
$$

\subsection{Wave functions and energy levels }

   We started by introducing the factorization (\ref{Psi(yLa)})
for the wave functions $\Psi$ and then, when the boundary
conditions have been taken into account, we have obtained that the first
factor $h(y,\La)$ must be a polynomial function of a very
particular class. The result is that the wave functions become
{\sl``a $\La$-dependent polynomial divided by the function $(1 +
\La\,y^2)^{1/(2\La)}$"}; more specifically,
$$
  \Psi_m(y,\La) = {\cal H}_m(y,\La)\,(1 + \La\,y^2)^{-\,1/(2\La)}
  \,,\quad m=0,1,2,\dots
$$
where $m$ is the order of the polynomial. Since the second factor
is even and has no nodes, the corresponding eigenfunction
$\Psi_m(y,\La)$ has the same parity as $m$ and  $m$ nodes
(when $\La<0$ the nodes are inside the the bounded interval
$[-\,a_\La,a_\La]$).
Therefore, these two particular properties, parity and number of
nodes, are preserved by the deformation introduced by  $\La$.

The two cases $\La$ negative and $\La$ positive are
rather different and must be studied separately.

(i) Let us suppose that $\La$ is negative, $\La<0$.

   In this case the dynamics is restricted to the interval
$y^2<1/|\La|$ and the boundary conditions imply that the function
$\Psi(y,\La)$ must vanish at the end points
$y_1=-\,1/\sqrt{|\La|}$ and $y_2= 1/\sqrt{|\La|}$.  This means
that the first factor $h(y,\La)$ must be a polynomial and, because
of this, the eigenfunctions are given by a countable family of functions
$\Psi_m$, $m = 0,1,2,\dots$.
The eigenfunctions $\Psi_m$ and the eigenvalues $e_m$ are given by
\begin{eqnarray}
  &&\Psi_m(y,\La) = {\cal H}_m(y,\La)\,(1 -
\left|\La\right|\,y^2)^{\,1/(2\left|\La\right|)}\,, \cr
   &&e_m = \left(m+\frac{1}{2}\right) + \frac{1}{2}\,m^2\,\left|\La\right|\,,
\qquad  m = 0,1,2,\dots,m, \dots
\end{eqnarray}
These eigenfunctions $\Psi_m$ are well defined and vanish at the
end points, $y_1$ and $y_2$, for all the integers values of $m$
without any restriction; thus, there exists a infinite but
countable number  of eigenfunctions. The energy spectrum is
unbounded but, contrarily to what happens in the case of the
linear oscillator, is not equispaced
\begin{eqnarray}
   &&e_0<e_1<e_2<e_3<\dots<e_m<e_{m+1}<\dots \cr&&\cr
   &&e_{m+1} - e_m = 1 + \left(m+\frac{1}{2}\right)\,\left|\La\right| \,.
\nonumber\end{eqnarray}
Nevertheless, it is also true that the energy $e_0$ of the
ground-state $\Psi_0$ is $e_0= 1/2$; that is, the zero-point
energy is given by $E_0=(1/2)\hbar\,\al$ for any $\La\ne 0$, as in the
linear case.

(ii) Let us suppose that $\La$ is positive, $\La>0$.

   This case is rather delicate and deserves to be analyzed with more detail.
It is convenient to recall that the domain of $y$ is now the whole
real line $\IR$ and hence it is necessary to take into account the
problem of the convergence at the infinity. In fact, it is
necessary that the following integral be convergent
$$
  \int_{-\infty}^{\infty} \frac{{\cal H}_m^2(y,\La)}
  {(1 + \La\,y^2)^{1/\La}\,\sqrt{1 + \La\,y^2}}\ dy\ <\ \infty
$$
and, as for large values of $y$, the powers of the dominant terms
in the numerator and the denominator are $2m$ and $1 + 2/\La$
respectively, we arrive to   a certain condition to be
satisfied by $m$. In fact, given a certain value of $\La$, then
the admissible functions $\Psi_m$ are those associated to
integer values of $m$ satisfying the condition
$$
  m < m_\La = \frac{1}{\La} \,.
$$
This means that for large values of $\La$ the oscillator only
admits the fundamental level $\Psi_0$ as eigenfunction, while for
smaller values of $\La$ the oscillator admits a certain finite number of
eigenstates, in such a way that when $\La$ decreases this number
increases and in the linear limit $\La\to 0$ the number goes to
infinity. The eigenfunctions $\Psi_m$ and the eigenvalues $e_m$
are given by
\begin{eqnarray}
  &&\Psi_m(y,\La) = {\cal H}_m(y,\La)\,(1 +\La\,y^2)^{-\,1/(2\La)}\,, \cr
  &&\cr
  &&e_m = \left(m+\frac{1}{2}\right) - \frac{1}{2}\,m^2\,\La\,,
\qquad   m= 0,1,2,\dots,N_\La,
\end{eqnarray}
where $N_\La$ denotes the greatest integer lower than $m_\La$. The
energy spectrum is bounded and again not equispaced
\begin{eqnarray}
   &&e_0<e_1<e_2<e_3<\dots<e_m<e_{N_\La}  \,,\cr&&\cr
   &&e_{m+1} - e_m = 1 - \left(m+\frac{1}{2}\right)\,\La \,.
\nonumber\end{eqnarray}
It can be seen that the energy $e_m$, considered as a function of
$m$, is an increasing function for low values of $m$ and has a
maximum at the point $1/\La$. Beyond that point it becomes a
decreasing function and it even takes negative values; but, due to
the above condition $m<m_\La$ for physical eigenfunctions, this
behaviour of $e_m$ for large values of $m$ is not physically
meaningful. Figure III shows the form of $e_m$ as a function of
$m$ for two particular values of $\La$.

The following particular cases, that are presented in a decreasing
way, illustrate the situation for several positive values of $\La$
\begin{itemize}
\item{} If $\La\ge 1$ the only bound state of this deformed
oscillator is the fundamental state $\Psi_0$.

\item{} If $1>\La\ge 1/2$  there are two eigenfunctions; the
fundamental level $\Psi_0$ and another bound state $\Psi_1$ with
energies $e_0<e_1$. For example, for $\La=0.8$ we have $e_0=1/2$
and $e_1=1.1$.

\item{} If $1/2>\La\ge 1/3$  there are three eigenfunctions
$\Psi_0$, $\Psi_1$, $\Psi_2$, with energies $e_0<e_1<e_2$. For
example, for $\La=0.4$ we have $e_0=1/2$, $e_1=1.3$, and $e_2 =
1.7$.

\item{} If $1/3>\La\ge 1/4$  there are four eigenfunctions
$\Psi_0$, $\Psi_1$, $\Psi_2$, $\Psi_3$, with energies
$e_0<e_1<e_2<e_3$.  For example, for $\La=0.3$ we have $e_0=1/2$,
$e_1=1.35$, $e_2=1.90$, and $e_3 = 2.15$.

\item{} If $1/(n-1)>\La\ge 1/n$ there exist $n$ eigenfunctions
(the fundamental one and $n-1$ other excited bound states)
$\Psi_0$ and $\Psi_i$, $i=1,2,\dots,n-1$, with energies $e_0=1/2
<e_1 <e_2<e_3<\dots<e_{n-1}$.
\end{itemize}

We close this section with a comparison of the energy levels
$e_m(\La)$ of the $\La$-dependent oscillator with the
corresponding values of the linear harmonic oscillator
\begin{eqnarray}
\left.\begin{array}{cl}
   & e_m(\La) = e_m(0) + \frac{1}{2}\,m^2\,\left|\La\right|\,,
   \quad \La<0 \cr&\cr
   & e_m(\La) = e_m(0) - \frac{1}{2}\,m^2\,\La\,, \qquad \La>0\ \cr
\end{array}\right\}
\end{eqnarray}
It is clear, therefore, that when $\La$ is negative the energy
$e_m(\La)$ is higher than the energy $e_m(0)$ of the harmonic oscillator
and when $\La>0$,  $e_m(\La)$, $m < m_\La$, is lower than $e_m(0)$.
Figure IV shows this property by plotting the values of
$e_m(\La)$ for $\La=-0.30$ and $\La=0.30$.

\section{Schr\"odinger factorization formalism }

The factorization of Schr\"odinger Hamiltonians in terms of first
order differential operators \cite{Sch40}-\cite{CaMaPR98} has been
shown to be very efficient in finding properties of the spectrum
of the Hamiltonian  and even exhaustively solving the spectral
problem in the case of shape invariant Hamiltonians
\cite{CaRa00Rmp}-\cite{CaRa00Jpa}.

In the usual case one starts with a Hamiltonian $H_1$ like
$$
  H_1 = -\,\frac{d^2}{dx^2} + V_1(x) \,,
$$
determined by the potential $V_1$ and looks for two operators $a$
and $a^+$ of the form
$$
  a  = \frac{d}{dx} + W(x)   \,,\qquad
  a^+ = -\,\frac{d}{dx} + W(x) \,,
$$
such that $H_1$ admits the factorization
\begin{equation}
  H_1 = a^+\,a =  \Bigl[\,-\,\frac{d}{dx} + W(x) \Bigr]
  \Bigl[\,\frac{d}{dx} + W(x) \Bigr]\,.
\end{equation}

This happens if and only if the function $W$ is such that
$$
  W' - W^2 + V_1 = 0 \,,
$$
i.e.,  $W$ is a solution of the preceding Riccati equation. When
$H_1$ admits such a factorization we can define a new Hamiltonian
$H_2$, of the form
$$
  H_2 = -\,\frac{d^2}{dx^2} + V_2(x)  \,,
$$
by means of the new factorization
\begin{equation}
  H_2 = a\,a^+ = \Bigl[\,\frac{d}{dx}+W(x) \Bigr]
  \Bigl[\,-\,\frac{d}{dx} + W(x) \Bigr]\,.
\end{equation}
Such a Hamiltonian $H_2$ is said to be the partner of $H_1$, and
leads to a new Riccati equation
$$
  W' + W^2 - V_2 = 0  \,.
$$
In summary, the function  $W$ is a solution of two different
Riccati equations, determined by the two  potentials $V_1$ and
   $V_2$:
\begin{eqnarray}
   &&W' - W^2 + V_1 = 0  \,,\cr
   &&W' + W^2 - V_2 = 0   \,,\nonumber
\end{eqnarray}
Therefore, the two  potentials $V_1$ and
   $V_2$ are related by
$$
  V_2 = 2 W^2 - V_1  \,.
$$
The important point is that $a$ is an intertwining operator for
both Hamiltonians and this property allows to establish relations
between the corresponding spectral problems.

  The method can be adapted to deal with the class of Hamiltonians we are
interested in:
$$
  \widehat{H}_1 = -\,(1 + \la\,x^2)\,\frac{d^2}{dx^2}
   - \la\,x\,\frac{d}{dx} + U_1(x)  \,,
$$
which correspond to a problem with a position dependent mass;
this is a particular example or factorization of Sturm--Liouville operators
\cite{HoSoA05}.
In this case we should look for two operators $A$ and $A^+$ of the 
following form
\begin{eqnarray}
   A  &=&  \sqrt{1 + \la\,x^2}\,\frac{d}{dx} + W(x) \,,\cr
  A^+ &=& -\,\sqrt{1 + \la\,x^2}\,\frac{d}{dx} + W(x)\,,\nonumber
\end{eqnarray}
and then, as
\begin{equation}
   \widehat{H}_1 = A^+\,A =
  \Bigl[\,-\,\sqrt{1 + \la\,x^2}\,\frac{d}{dx} + W(x)\, \Bigr]
  \Bigl[\,\sqrt{1 + \la\,x^2}\,\frac{d}{dx} + W(x)\, \Bigr]
\end{equation}
such a factorization is possible if and only if the function $W$ satisfies
the Riccati differential equation
$$
   \sqrt{1 + \la\,x^2}\,W' - W^2 + U_1 = 0.
$$
The partner Hamiltonian is now of the form
$$
  \widehat{H}_2 = -\,(1 + \la\,x^2)\,\frac{d^2}{dx^2}
  -\,\la\,x\,\frac{d}{dx} + U_2(x)  \,,
$$
and is defined by the alternative factorization
\begin{equation}
   \widehat{H}_2 = A\,A^+ =
   \Bigl[\,\sqrt{1 + \la\,x^2}\,\frac{d}{dx} + W(x)\, \Bigr]
   \Bigl[\,-\,\sqrt{1 + \la\,x^2}\,\frac{d}{dx} + W(x)\, \Bigr]
\end{equation}
which leads to
$$
  \sqrt{1 + \la\,x^2}\,W' + W^2 - U_2 = 0.
$$

In other words, the function  $W$ defining the factorization should be a
solution of two Riccati equations defined by the two potential functions
$U_1$ and  $U_2$
\begin{eqnarray}
  &&\sqrt{1 + \la\,x^2}\,W' - W^2 + U_1 = 0  \,,\cr
  &&\sqrt{1 + \la\,x^2}\,W' + W^2 - U_2 = 0  \,,\nonumber
\end{eqnarray}
and therefore both potentials are related by
$$
  U_2 = 2 W^2 - U_1  \,.
$$

\section{Schr\"odinger factorization approach to the $\la$-depen\-dent
nonlinear oscillator }

In a recent paper \cite{CaRaS04} a factorization for the
$\la$-dependent quantum  nonlinear oscillator was proposed and it
was shown to have the shape-invariance property, therefore the
spectrum of such system  has been fully determined. We aim in this
section to do a more complete computation of both the spectrum and
the corresponding eigenfunctions and to study some interesting
properties of the polynomial replacing Hermite polynomials in this
deformed case.

\subsection{Operators $A$ and $A^+$ and intertwined Hamiltonians
$H_1$ and $H_2$}

It has been proved in \cite{CaRaS04}  that  the following
operators $A$ and $A^+$
\begin{eqnarray}
  A &=& \frac{\hbar}{\sqrt{2 m}}\,\Bigl[\,\sqrt{1 + \la\,x^2}\,
  \frac{d}{dx} + (\frac{m\,\al}{\hbar})\,\frac{x}{\sqrt{1 + 
\la\,x^2}}\,\Bigr]  \,,\cr
  A^+ &=& \frac{\hbar}{\sqrt{2 m}}\,\Bigl[\,-\,\sqrt{1 + \la\,x^2}\,
  \frac{d}{dx} + (\frac{m\,\al}{\hbar})\,\frac{x}{\sqrt{1 + \la\,x^2}}\,\Bigr]
  \,,\nonumber
\end{eqnarray}
provide a factorization for the $\la$-dependent quantum  nonlinear
oscillator. In fact, if we denote by $W_{\la}$ the following function
$$
  W_{\la} = \frac{x}{\sqrt{1 + \la\,x^2}}\,,
$$
then the product $A^+\,A$ that is given by
$$
  A^+\,A = \frac{\hbar^2}{2 m}\,\Bigl[\,-\,\sqrt{1 + \la\,x^2}\,
  \frac{d}{dx} + (\frac{m\,\al}{\hbar})\,W_{\la} \,\Bigr] \Bigl[\,\sqrt{1 +
  \la\,x^2}\,\frac{d}{dx} + (\frac{m\,\al}{\hbar})\,W_{\la} \,\Bigr] \,,
$$
turns out to be
\begin{eqnarray}
  A^+\,A &=& -\,\frac{\hbar^2}{2 m}\,\Bigl[\,(1 + \la\,x^2)\,
  \frac{d^2}{dx^2} + \la\,x\,\frac{d}{dx}\,\Bigr] + U_1  \,,\cr
  U_1 &=& \frac{1}{2}\,(m\,\al)\bigl(\al + \frac{\hbar}{m}\,\la\bigr)
  \,W_{\la}^2 - \frac{1}{2}\,(\hbar\al)   \,.
\end{eqnarray}
In a similar way, the  product $A\,A^+$ given by
$$
   A\,A^+ = \frac{\hbar^2}{2 m}\,\Bigl[\,\sqrt{1 + \la\,x^2}\,\frac{d}{dx}
   + (\frac{m\,\al}{\hbar})\,W_{\la} \,\Bigr] \Bigl[\,-\,\sqrt{1 + \la\,x^2}\,
   \frac{d}{dx} + (\frac{m\,\al}{\hbar})\,W_{\la} \,\Bigr] \,,
$$
turns out to be
\begin{eqnarray}
  A\,A^+ &=&   -\,\frac{\hbar^2}{2 m}\,\Bigl[\,(1 + \la\,x^2)\,
  \frac{d^2}{dx^2} + \la\,x\,\frac{d}{dx}\,\Bigr] + U_2  \,,\cr
  U_2 &=& \frac{1}{2}\,(m\,\al)\bigl(\al - \frac{\hbar}{m}\,\la\bigr)
  \,W_{\la}^2 + \frac{1}{2}\,(\hbar\al)    \,.
\end{eqnarray}

Consequently, the Hamiltonian $\widehat{H}$ given by
(\ref{Hq(xla)}), is such that
$$
  \widehat{H} = \widehat{H}_1+\frac{1}{2}\,(\hbar\al)
  = \widehat{H}_2-\frac{1}{2}\,(\hbar\al)
$$
where $\widehat{H}_1$ and $\widehat{H}_2$ are  the Hamiltonian
$\widehat{H}_1 = A^+\,A$  and its partner $\widehat{H}_2= A\,A^+$.

Note that the limits when $\la\to 0$ are given by:
\begin{eqnarray}
   &&\lim{}_{\la{\to}0}\,\widehat{H}_1 = - \frac{\hbar^2}{2 m}\,\frac{d^2}{dx^2}
   + (\frac{1}{2})\,m\,\al^2x^2 - \bigl(\frac{1}{2}\bigr)(\hbar\al) \,,\cr&&\cr
   &&\lim{}_{\la{\to}0}\,\widehat{H}_2 = - \frac{\hbar^2}{2 m}\,\frac{d^2}{dx^2}
   + (\frac{1}{2})\,m\,\al^2x^2 + \bigl(\frac{1}{2}\bigr)(\hbar\al) \,.\nonumber
\end{eqnarray}
and note also that the  commutator $[A,A^+]$ is not a  constant. In fact,
$$
  [\,A\,,\,A^+\,] = \left(1 -\,\frac{\la\,x^2}{1 + \la\,x^2}\right)
(\hbar\,\alpha)
$$
but when $\la\to 0$ we find
$$
  \lim{}_{\la{\to}0}\,[\,A\,,\,A^+\,] = \hbar\,\al  \,.
$$

\subsection{The  spectrum of the quantum deformed nonlinear oscillator}

  The two operators, the Hamiltonian $\widehat{H}_1$ and its partner
$\widehat{H}_2$, depend on the parameter $\alpha$; the important
point is that this $\alpha$-dependence determines an important
relation between them given by
\begin{equation}
  \widehat{H}_2(\al) = \widehat{H}_1(\al_1) + R(\al_1)  \,,\label{shinvcon}
\end{equation}
where $\alpha_1$ denotes $\alpha_1=f(\alpha)$ with the functions $f$ and $R$
defined by the expressions
\begin{equation}
  f(\al) = \al-\frac{\hbar}{m}\,\la  \,,\qquad
  R(\al) = \hbar \al + \bigl(\frac{1}{2}\bigr)\frac{\hbar^2}{m}\la \,.
\label{Def(fR)}
\end{equation}

  The condition (\ref{shinvcon}) is called `shape invariance'
condition and for such cases there exists a recipe to compute the
spectrum and to determine the corresponding eigenfunctions
\cite{Ge83,GeKr85} (see also \cite{CaRa00Rmp} for a modern approach).

  More specifically, as a first step, the bound state $\ket{\Psi_0(\alpha)}$
is found by solving $A(\alpha)\ket{\Psi_0(\alpha)}=0$, and has a
zero energy, i.e. $\widehat{H}_1(\alpha)\ket{\Psi_0(\alpha)}=0$.
Then, using (\ref{shinvcon}) we can see that
$\ket{\Psi_0(\alpha_1)}$ is an eigenstate of
$\widehat{H}_2(\alpha)$ with an  energy $E_1=R(\alpha_1)$, because
\begin{equation}
  \widehat{H}_2(\alpha)\ket{\Psi_0(\alpha_1)}
  = (\widehat{H}_1(\alpha_1)+R(\alpha_1))\ket{\Psi_0(\alpha_1)}
  = R(\alpha_1)\ket{\Psi_0(\alpha_1)}\,.
\end{equation}

  Next, $A^{\dag}(\alpha)\ket{\Psi_0(\alpha_1)}$ is an eigenstate of
$\widehat{H}_1(\alpha)$ with an  energy $E_1=R(\alpha_1)$, because
if we use the intertwining relation
$\widehat{H}_1(\alpha)A^{\dag}(\alpha) =
A^{\dag}(\alpha)\widehat{H}_2(\alpha)$ we see that
$$
  \widehat{H}_1(\alpha)A^{\dag}(\alpha)\ket{\Psi_0(\alpha_1)}
  = A^{\dag}(\alpha)\widehat H_2(\alpha)\ket{\Psi_0(\alpha_1)}
  = A^{\dag}(\alpha)(\widehat{H}_1(\alpha_1) + R(\alpha_1))
  \ket{\Psi_0(\alpha_1)}  \,,
$$
and hence we arrive to
$$
  \widehat{H}_1(\alpha)A^{\dag}(\alpha)\ket{\Psi_0(\alpha_1)}
  =  R(\alpha_1)A^{\dag}(\alpha)\ket{\Psi_0(\alpha_1)}\,.\nonumber
$$
This process should be iterated and we will find the sequence of
energies for $\widehat{H}_1(\alpha)$
\begin{equation}
  E_k = \sum_{j=1}^k R(\alpha_j)\,,\qquad  E_0=0\,,
\end{equation}
the corresponding eigenfunctions being
\begin{equation}
  \ket{\Psi_n(\alpha_0)}=A^{\dag}(\alpha_0)A^{\dag}(\alpha_1)\cdots
  A^{\dag}(\alpha_{n-1})\ket{\Psi_0(\alpha_n)}\,,
\end{equation}
where $\alpha_0=\alpha$ and  $\alpha_{j}=f(\alpha_{j-1})$, namely,
$\alpha_j=f^j(\alpha_0)=f^j(\alpha)$.

  We can now apply this procedure with the functions $f$ and $R$
given by (\ref{Def(fR)}). In this case we obtain
\begin{eqnarray}
&&\alpha_k=f^k(\alpha)= \alpha-\frac{\hbar\lambda}m\,k  \,,\cr
&&R(\alpha_k)= h\alpha_k+ \frac{\hbar^2}{2\,m}\,
  \lambda= \hbar\Bigl(\alpha-\frac{\hbar\lambda}m\,k\Bigr) +
  \frac{\hbar^2}{2 m}\,\lambda
\end{eqnarray}
and therefore the energy $E_n$ that is given by
$$
  E_n = \sum_{k=1}^n R(\alpha_k)  = \sum_{k=1}^n
  \left(\hbar\Bigl(\alpha-\frac{\hbar\lambda}m\,k\Bigr)
  + \frac{\hbar^2}{2 m}\lambda\right)
$$
becomes
$$
  E_n = \left(n\,\alpha-\frac{\hbar\lambda}{2m}\,n(n+1)\right)\hbar
  + \frac{n\,\hbar^2 }{2\,m} \lambda
  = \Bigl(n\,\alpha-\frac{n^2\,\hbar\,\lambda}{2\,m}\Bigr) \hbar\,,
$$
that in terms of the adimensional variable $\La$ becomes
\begin{equation}
  E_n = \Bigl[\,n - (\frac{1}{2})\,n^2\,\La \,\Bigr](\hbar\,\al)
  \,,\quad n=0,1,2,\dots
\end{equation}
The eigenstates of $\widehat{H}$ will include an additional term
$(1/2)\hbar\alpha$ which shifts all levels.

\subsection{Eigenfunctions of the quantum deformed nonlinear oscillator}

  As indicated before, the fundamental state is determined by
$A\ket{\Psi_0 }=0$. The wave function $\Psi_0(x)$ in the
coordinate representation is then given by the normalized solution
of the linear first order equation:
$$
  (1 + \la\,x^2)\,\frac{d}{dx}\,\Psi_0 + \be\,x\,\Psi_0 = 0
  \,,\qquad\be = \frac{m\,\al}{\hbar} \,,
$$
whose solution is proportional to
$$
  \Psi_0 = \frac{1}{(1 + \la\,x^2)^{\,r_0}} \,,\quad
  r_0 = \frac{\be}{2\la} = \frac{1}{2\La}\,.
$$

In the limit of $\lambda$ going to zero, having in mind that
$\lim_{\lambda\to  0}(1+\la\, x^2)=1$ and $\lim_{\lambda\to 0}
\beta/(2\,\lambda)= \infty$, we have that
$$
  \lim_{\lambda\to 0} (1 + \la\,x^2)^{-\be/2\la} =
  \exp\left(-\lim_{\lambda\to 0} \frac{\log (1 +\la\,x^2)}
  {2\lambda/\beta}\right)= \exp\left(-\frac{\beta}{2}\,x^2\right)\,.
$$

The other eigenfunctions are obtained according to the recipe we
derived before as follows:
\begin{eqnarray}
  \Psi_1 &=& A^+(\al_0)\,\Psi_0(\al_1) \cr
  \Psi_2 &=& A^+(\al_0)\,A^+(\al_1)\,\Psi_0(\al_2) \cr
  \Psi_3 &=& A^+(\al_0)\,A^+(\al_1)\,A^+(\al_2)\,\Psi_0(\al_3) \cr
  \dots& & \dots \dots   \cr
  \Psi_n &=& A^+(\al_0)\,A^+(\al_1)\,A^+(\al_2)\,\,\dots\,A^+(\al_{n-1})\,
  \Psi_0(\al_n)
\end{eqnarray}

  In the next property we establish a powerful tool for computing the
action of different operators $A^+(\al_k)$ on functions
$\Psi_0(\al_j)$.

\begin{proposicion}
Let $g(x)$ be an arbitrary differentiable function. Then the following
equality holds:
$$
  (1 + \la\,x^2)^p\,\sqrt{1 + \la\,x^2}\,\frac{d}{dx}\,(1 
+\la\,x^2)^{-\,p}\,g(x)
  = (-1) \Bigl[-\,\sqrt{1 + \la\,x^2}\,\frac{d}{dx}
  + \frac{(2 p \la)\,x}{\sqrt{1 + \la\,x^2}} \,\Bigr] g(x) \,.
$$
\end{proposicion}
{\it Proof:} This property is proved by direct computation.

Using the notation $z_x= 1 + \la\,x^2$, the above
property can be rewritten as follows
$$
  z_x^p\,\sqrt{z_x}\,\frac{d}{dx}\,z_x^{-\,p}
  = (-1)\left[-\,\sqrt{z_x}\,\frac{d}{dx} + \frac{(2 p\la)\,x}
  {\sqrt{z_x}}\,\right]   \,,
$$
in such a way that in the particular case $p=b/2\la$ it reduces to
$$
  z_x^{b/2\la}\,\sqrt{z_x}\,\frac{d}{dx}\,z_x^{-\,b/2\la}
  = (-1)\left[-\,\sqrt{z_x}\,\frac{d}{dx} + \frac{b\,x}{\sqrt{z_x}}\,\right] \,.
$$

The operators  $A^+(\al_0)$, $A^+(\al_1)$, $A^+(\al_2)$, $\dots$,
which are given by
\begin{eqnarray}
   A^+(\al_0) &=& C_0\,\bigl[-\,\sqrt{z_x}\,\frac{d}{dx}
+(\frac{m\,\al_0}{\hbar})\frac{x}{\sqrt{z_x}} \,\bigr] \,, \cr
   A^+(\al_1) &=& C_1\,\bigl[-\,\sqrt{z_x}\,\frac{d}{dx}
+(\frac{m\,\al_1}{\hbar})\frac{x}{\sqrt{z_x}} \,\bigr] \,, \cr
   A^+(\al_2) &=& C_2\,\bigl[-\,\sqrt{z_x}\,\frac{d}{dx}
+(\frac{m\,\al_2}{\hbar})\frac{x}{\sqrt{z_x}} \,\bigr] \,, \cr
\dots& & \dots \dots \dots    \nonumber
\end{eqnarray}
can be rewritten as follows
\begin{eqnarray}
  A^+(\al_0) &=& C_0\,(-1)\,
  z_x^{\be_0/2\la}\,\sqrt{z_x}\,\frac{d}{dx}\,z_x^{-\,\be_0/2\la} \,,\cr
  A^+(\al_1) &=& C_1\,(-1)\,
  z_x^{\be_1/2\la}\,\sqrt{z_x}\,\frac{d}{dx}\,z_x^{-\,\be_1/2\la} \,,\cr
  A^+(\al_2) &=& C_2\,(-1)\,
  z_x^{\be_2/2\la}\,\sqrt{z_x}\,\frac{d}{dx}\,z_x^{-\,\be_2/2\la} \,,\cr
\dots& & \dots \dots \dots  \nonumber
\end{eqnarray}
where $\be_0$ and $\be_k$ denote $\be_0=\be$ and $\be_k=(m\,\al_k)/\hbar$.

Finally, the functions  $\Psi_0$, $\Psi_1$, $\Psi_2$, $\Psi_3$,
$\dots$, $\Psi_n$, $\dots$,  turn out to be
\begin{eqnarray}
  \Psi_0 &=& C_0\,z_x^{-\,\be/2\la} \,, \cr
  \Psi_1 &=& C_0C_1\,(-1)\,
  z_x^{-\,\be/2\la}\,\bigl[\,z_x^{\be/\la}\,\cdot\,\sqrt{z_x}\,
   \frac{d}{dx}\,(\sqrt{z})\,z_x^{-\,\be_0/\la}\,\bigr]  \,, \cr
  \Psi_2 &=& (C_0\dots C_2)\,
  (-1)^2\,z_x^{-\,\be/2\la}\,\bigl[\,z^{\be/\la}\,\cdot\,\sqrt{z_x}\,
   \frac{d^2}{dx^2}\,(\sqrt{z_x})^3\,z_x^{-\,\be_0/\la}\,\bigr] \,, \cr
  \Psi_3 &=&  (C_0\dots C_3)\,
  (-1)^3\,z_x^{-\,\be/2\la}\,\bigl[\,z_x^{\be/\la}\,\cdot\,\sqrt{\,z\,}\,
  \frac{d^3}{dx^3}\,(\sqrt{z_x})^5\,z_x^{-\,\be_0/\la}\,\bigr]  \,, \cr
  \dots& & \dots \dots \dots \dots  \nonumber
\end{eqnarray}
and therefore the $n$-th wave-function
$\Psi_n(x) $ is given by
\begin{equation}
   \Psi_n = (C_0\dots C_n)\,(-1)^n\,\Bigl[\,z_x^{\be/\la}\,\cdot\,\sqrt{z_x}\,
   \frac{d^n}{dx^n}\,(\sqrt{z_x})^{2n-1}\,z_x^{-\,\be_0/\la}\,\Bigr]\,
z_x^{-\,\be/2\la} \label{Psin(xla)}
\end{equation}
that must be considered as a new way of representing the
expression (\ref{Psi(yLa)}) obtained in the Sec. II by the direct
approach to the differential equation.

\section{$\la$-dependent Hermite polynomials }

   We have solved the $\la$-dependent nonlinear oscillator by using
two different procedures and in both cases we have arrived to an
expression for the wave-function $\Psi_n(x)$ depending of some
$\la$-dependent polynomials that must be considered as a
deformation of the Hermite polynomials.
A natural idea is to expect that the classical properties of the
Hermite polynomials must remain true,
with the appropriate modifications, for these new $\la$-dependent
Hermite polynomials.  In Sec. II we proved the orthogonality, now
we will study three other fundamental properties: $\la$-dependent
Rodrigues formula, the existence of a $\la$-dependent generating
function and the existence of recursion relations among polynomials
of different orders.

The Schr\"odinger factorization approach to the $\la$-dependent
nonlinear oscillator, studied in Sec. IV, has provided the
expression (\ref{Psin(xla)}) for the wave function $\Psi_n$.
Therefore, we have
$$
  {\cal H}_n(x,\la) = (-1)^n\,z_x^{\be/\la+1/2}\,
  \frac{d^n}{dx^n}\,\Bigl[\,z_x^{n}\,z_x^{-\,(\be/\la+1/2)}\,\Bigr]
  \,,\quad n=0,1,2,\dots
$$
or
\begin{equation}
  {\cal H}_n(y,\La) = (-1)^n\,z_y^{1/\La+1/2}\,
  \frac{d^n}{dy^n}\,\Bigl[\,z_y^{n}\,z_y^{-\,(1/\La+1/2)}\,\Bigr]
  \,,\quad z_y = 1 + \La\,y^2  \,,
\end{equation}
that must be considered as the ``Rodrigues formula'' for the
$\la$-dependent Hermite polynomials obtained in Sec. II. In fact,
it is clear that
$$
  \lim{}_{\La{\to}0}\,\biggl[\, (-1)^n\,z_y^{1/\La+1/2}\,
  \frac{d^n}{dy^n}\,\Bigl[\,z_y^{n}\,z_y^{-\,(1/\La+1/2)}\,\Bigr]\,\biggr]
  = (-1)^n\,e^{y}\,\frac{d^n}{dy^n}\,e^{-y}
$$
and consequently
$$
  \lim{}_{\La{\to}0}\,{\cal H}_n(y,\La) = H_n(y)
  \,,\quad n=0,1,2,\dots
$$
  In a similar way to the standard Hermite polynomials, every function
${\cal H}_m$ is a  polynomial of degree $m$ and it has a well defined
parity
$$
  {\cal H}_m = c_m y^m + c_{m-2} y^{m-2} + c_{m-4} y^{m-4} +\dots
$$
in such a way that the coefficient $c_m$ of $x^m$ in ${\cal H}_m$
is given by
$$
  c_m = \prod\nolimits_{r=m}^{2m-1}\,(2 - r\,\La)
  = (2 - m\,\La)(2 - (m+1)\,\La) \dots (2 - (2m-1)\,\La)
$$
so that
$$
  \lim\nolimits_{\La\to 0} c_m = 2^m  \,.
$$
The first polynomials ${\cal H}_n={\cal H}_n(y,\La)$,
$n=0,1,2,\dots,6$, have the following expressions
\begin{eqnarray}
   {\cal H}_0 &=& 1  \cr
   {\cal H}_1 &=& k_1\,y  \cr
   {\cal H}_2 &=& k_2\,[\,2(1 - \La)y^2 - 1\,] \cr
   {\cal H}_3 &=& k_3\,[\,2(1 - 2\La)y^3 - 3y \,]    \cr
   {\cal H}_4 &=& k_4\,[\,4(1 - 2\La)(1 - 3\La)y^4 - 12(1 - 2\La)y^2
    + 3\,] \cr
   {\cal H}_5 &=& k_5\,[\,4(1 - 3\La)(1 - 4\La)y^5 - 20(1 - 3\La)y^3
    + 15y \,] \cr
   {\cal H}_6 &=& k_6\,[\,8(1 - 3\La)(1 - 4\La)(1 - 5\La)y^6
   - 60(1 - 3\La)(1 - 4\La)y^4 + 90(1 - 3\La)y^2 - 15\,]
\nonumber\end{eqnarray}
where the constants $k_i$, $i=1,2,\dots,6$, are given by
\begin{eqnarray}
  k_1 &=& (2 - \La)  \,,\hskip108pt
  k_2 = (2 - 3\La)  \,,\cr
  k_3 &=& (2 - 3\La)(2 - 5\La) \,,\hskip62pt
  k_4 = (2 - 5\La)(2 - 7\La)    \,,\cr
  k_5 &=& (2 - 5\La)(2 - 7\La)(2 - 9\La) \,,\qquad
  k_6 = (2 - 7\La)(2 - 9\La)(2 - 11\La) \,.\nonumber
\end{eqnarray}

We next analyze the existence of a generating function.

A generating function for the $\La$-dependent Hermite polynomials
${\cal H}_n(y,\La)$ must be a function ${\cal F}(t,y,\La)$ that
generates such polynomials as the successive coefficients of its
Taylor  power series of the additional variable $t$. Since the
generating function of the Hermite polynomials is known to be
given by
$$
  F(t,y) = \sum\nolimits_{n=0}^{\infty} (\frac{1}{n\,!})\,H_n(y)\,t^n  \,,\quad
  F(t,y) = e^{(2 t y - t^2)} \,,
$$
then, a necessary condition to be satisfied by ${\cal F}(t,y,\La)$
must be the fulfillment of the limit
$$
  \lim{}_{\La{\to}0}\,{\cal F}(t,y,\La) = e^{(2 t y - t^2)} \,.
$$
We propose to consider  the function
\begin{equation}
  {\cal F}(t,y,\La) = \Bigl(1 + \La\,(2 t y - t^2)\Bigr)^{1/\La}
\end{equation}
as the appropriate function for representing the $\La$-dependent
generating function. It is clear that it satisfies the correct limit
$$
  \lim{}_{\La{\to}0}\,\Bigl(1 + \La\,(2 t y - t^2)\Bigr)^{1/\La}
   = e^{(2 t y - t^2)}   \,.
$$

The power series of ${\cal F}$ in the new variable $t$ is given by
\begin{equation}
  \Bigl(1 + \La\,(2 t y - t^2)\,\Bigr)^{1/\La}
   = \sum\nolimits_{n=0}^{\infty} (\frac{1}{n\,!})\,\wt{\caH}_n(y,\La)\,t^n
\end{equation}
where we have used the notation $\wt{\caH}_n$ for the coefficient
of the Taylor series. The first polynomials $\wt{\caH}_n= \wt{\caH}_n(y,\La)$,
$n=0,1,2,\dots,6$, obtained in such a way, have the
following expressions
\begin{eqnarray}
  \wt{\caH}_0 &=& 1  \cr
  \wt{\caH}_1 &=& 2\,g_1\,y  \cr
  \wt{\caH}_2 &=& 2\,g_2\,[\,2(1 - \La)y^2 - 1\,]  \cr
  \wt{\caH}_3 &=& 4\,g_3\,[\,2(1 - 2\La)y^3 - 3 y\,]    \cr
  \wt{\caH}_4 &=& 4\,g_4\,[\,4(1 - 2\La)(1 - 3\La)y^4 - 12(1 - 2\La)y^2
  + 3\,]\cr
  \wt{\caH}_5 &=& 8\,g_5\,[\,4(1 - 3\La)(1 - 4\La)y^5 - 20(1 -3\La)y^3
  + 15\,y\,]\cr
  \wt{\caH}_6 &=& 8\,g_6\,[\,8(1 - 3\La)(1 - 4\La)(1 - 5\La)y^6
  - 60(1 - 3\La)(1 - 4\La)y^4 + 90(1 - 3\La)y^2 - 15\,]
\nonumber\end{eqnarray}
where the constants $g_i$, $i=1,2,\dots,6$, take the values
$$
  g_1 = g_2 = 1\,,\quad g_3 = g_4 = (1-\La)\,,\quad
  g_5 = g_6 = (1-\La)(1-2\La) \,.
$$

Hence, the polynomials $\wt{\caH}_n$, obtained from the generating
function ${\cal F}(t,y,\La)$, are essentially the same ones that
the polynomials ${\cal H}_n$, obtained from the Rodrigues formula.
They coincide in the fundamental part ($y$-dependent polynomial part
written between square brackets [\dots]) and only differ in the values
of the global multiplicative coefficients that now are given by $g_i$ instead
of $k_i$
(although $g_i$ and $k_i$ are rather similar nevertheless $g_i\ne k_i$).

The existence of this generating function allows us to establish
the existence of $\La$-dependent recursion relations among
deformed Hermite polynomials of different orders.

Firstly, let us take the derivative of ${\cal F}(t,y,\La)$ with
respect to $t$ and then multiply both sides by $1 + \La\,(2 t y -
t^2)$. In this way we obtain the relation
$$
  2(y-t) \Biggl(\,\sum_{k=0}^{\infty}
  \frac{1}{k\,!}\,\wt{\caH}_k(y,\La)\,t^k\Biggr)
  = \Bigl(1 + \La\,(2 t y - t^2)\Bigr) \Biggl(\,\sum_{k=0}^\infty
  \frac{1}{k!}\,\wt{\caH}_{k+1}(y,\La)\,t^k\Biggr)
$$
that leads to
$$
  \wt{\caH}_1 = 2\,y\,\wt{\caH}_0
$$
and to the following recursion relation
$$
  \wt{\caH}_{n+1} = 2\,y\,(1- n\,\La) \wt{\caH}_{n}
  - n\,(2 - (n-1)\La)\wt{\caH}_{n-1} \,,\quad n\geq 1 \,.
$$
It is clear that if we particularize this formula for $\La=0$
then we recover the recursion formula of the standard Hermite
polynomials.

  Finally, if we take the derivative of the function ${\cal F}(t,y,\La)$ with
respect to the variable $y$ and then multiply both sides by $1 + 
\La\,(2 t y - t^2)$
we arrive to the following relation
$$
  2 t \,\Bigl(1 + \La\,(2 t y - t^2)\Bigr)^{1/\La}
  = \Bigl(1 + \La\,(2 t y - t^2)\Bigr) \Biggl(\,\sum_{k=0}^\infty
  \frac{1}{k!}\,\wt{\caH}_k^{\,'}(y,\La)\,t^k\Biggr)
$$
so that, equating the coefficients of each power of $t$, we obtain
\begin{eqnarray}
  &&\wt{\caH}_0^{\,'} = 0  \cr
  &&\wt{\caH}_2^{\,'} + 4\La\,y\,\wt{\caH}_1^{\,'} - 2\La\,y\,\wt{\caH}_0^{\,'}
  = 2 \wt{\caH}_0
\nonumber\end{eqnarray}
as well as the general expression
$$
  \wt{\caH}_{n+2}^{\,'} + (n+2)\La\,\Bigl[\,2\,y\,\wt{\caH}_{n+1}^{\,'}
  - (n+1) \wt{\caH}_n^{\,'}\,\Bigr] = 2 (n+2)\wt{\caH}_{n+1}
  \,,{\quad} n=0,1,2,\dots
$$
Once more we obtain for $\La=0$ a well known property of the standard Hermite
polynomials.

\section{Final comments and outlook}

   We have have studied the quantum version of the $\la$-dependent
one-dimensional non-linear oscillator of Mathews and Lakshmanan.
The first dificulty to deal with was the question of the order
ambiguity in the quantization of the Hamiltonian since it is a system
with a position-dependent mass; this problem was solved by introducing a
prescription (previously studied in Ref. \cite{CaRaS04}) obtained
from the analysis of the properties of the classical system
(existence of a Killing vector and a $\la$-dependent invariant
measure). The quantum Hamiltonian obtained by this procedure has
proved to belong to the restricted family of exactly solvable
Hamiltonians.
This is a very interesting situation: quantum systems with
nonpolynomial rational potentials (and standard kinetical term)
are hard to study and difficult to be solved and, on the other
hand, quantum systems with a position-dependent mass are also
difficult to study and generally not exactly solvable.
The system we have considered combines however both traits
in a very specific way, and this combination turns the problem
exactly solvable.

  We recall that has been proved in \cite{CaRaSS04} that although the
classical two-dimensional nonlinear oscillator is not separable in
cartesian coordinates it admits however Hamilton-Jacobi separability
in three other different coordinate systems (polar coordinates and
two other $\la$-dependent systems).
It is known that classical Hamilton-Jacobi separability and quantum
Schr\"odinger separability are closely related, so it is to be expected
that the $n=2$ quantum system will also admit a solvable Schr\"odinger
equation.
We think that this question must be investigated, as well as its
connections to other approaches related with the factorization of
two-dimensional systems \cite{KuTeV01},\cite{SaZa05},\cite{EfSp97}.
For instance, it was shown in \cite{EfSp97} that the Calogero and
Calogero--Sutherland models \cite{Ca69,Su71} have the
generalized shape invariance property.

   It is intresting to remark the existence of another family of
Hermite-related polynomials connected with the harmonic oscillator:
the so called ``relativistic Hermite polynomials" introduced in
\cite{AlBiN91} in the study of quantum relativistic
harmonic oscillators (see also \cite{AlBiGu96} and \cite{AlGu05}
and references therein).
Although our approach has been entirely non-relativistic and
the ``$\lambda$-deformed Hermite" polynomials ${\caH}_n$ (or $\wt{\caH}_n$)
seem to be rather different to these ``relativistic Hermite polynomials",
the existence of some kind of relationship seems quite probable and would
be worth studying.

  Finally, also in Ref. \cite{CaRaSS04}, a geometrical interpretation
of this nonlinear system was proposed (a similar interpretation is
discussed in \cite{BoDaK93,BoDaK94} with $\la$ in the kinetic
term and a $\la$-independent potential).
The main idea is that this $\la$-dependent system
represents a nonlinear model for the harmonic oscillator in the
circle $S^1$ ($\la<0$) or in the hyperbolic line $H^1$ ($\la>0$).
Remark that in the $n=1$ case, it is convenient to consider $S^1$
and $H^1$ as one-dimensional spaces obtained by endowing each
single geodesic of $S^2$ or $H^2$ with the induced metric (motion
on $S^1$ and $H^1$ will correspond to the $J=0$ radial motions on
$S^2$ or $H^2$). The point is that, if we assume that this
geometric interpretation is correct, then the $\la$-dependent
``deformed Hermite" polynomials ${\caH}_n$ (or $\wt{\caH}_n$)
could be interpreted as the ``curved" version of the Hermite
polynomials on the circle $S^1$  or in the hyperbolic line $H^1$.
We think that this is another question to be studied.

\section*{\bf Acknowledgments.}
Support of projects  BFM-2003-02532, FPA-2003-02948, E23/1 (DGA),
MTM-2005-09183, and VA-013C05 is acknowledged.

{\small

 }

\vfill\eject
\section*{\bf Figure Captions}

\begin{itemize}

\item{} {\sc Figure I}.{\enskip}
Plot of $V_\la(x)=(1/2)\,(\al^2 x^2)/(1 + \la\,x^2)$, $\al=1$, $\la<0$,
as a function of $x$, for $\la=-2$
(upper curve), and $\la=-1$ (lower curve).

\item{} {\sc Figure II}.{\enskip}
Plot of $V_\la(x)=(1/2)\,(\al^2 x^2)/(1 + \la\,x^2)$, $\al=1$, $\la>0$,
as a function of $x$, for $\la=1$
(upper curve), and $\la=2$ (lower curve).

\item{} {\sc Figure III}.{\enskip}
Plot of the energy $e_m$ as a function of $m$ for $\La=0.30$
(lower curve) and $\La=0.15$ (upper curve).
The curves also show the plot of the points $(m,e_m)$ for the
values $m=0, 1, 2, 3$, and $m=0, 1, \dots, 6$, respectively. For
$\La=0.30$ there exist four bound states (four thick points in the
curve) and for $\La=0.15$ seven bound states (seven thick points
in the curve). When $\La$ decreases the maximum of the curve moves
into the up right and in the limit $\La\to 0$ the curve converges
into a straight line parallel to the diagonal (dashed line).

\item{} {\sc Figure IV}.{\enskip}
Plot of the energy $e_m$ as a function of $m$ for $\La=0.30$
(lower curve) and  $\La=-0.30$ (upper curve).
The thick points $(m,e_m)$, corresponding to the values
$m=0, 1, 2, 3$, represent the four bound states existing
for $\La=0.30$ and the first four bound states for $\La=-0.30$.
The straight line (dashed line) placed in the middle corresponds
to the linear harmonic
oscillator.

\end{itemize}

\vfill\eject

$$
    \epsfbox{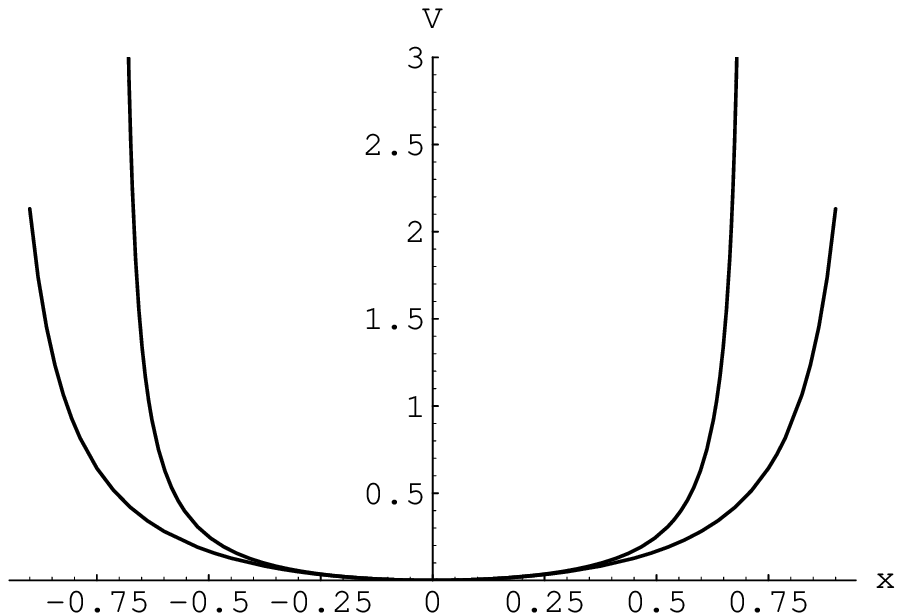}
$$

{\sc Figure I}.{\enskip} {\enskip}
Plot of $V_\la(x)=(1/2)\,(\al^2 x^2)/(1 + \la\,x^2)$, $\al=1$, $\la<0$,
as a function of $x$, for $\la=-2$
(upper curve), and $\la=-1$ (lower curve).

{\vskip 40pt}
$$
    \epsfbox{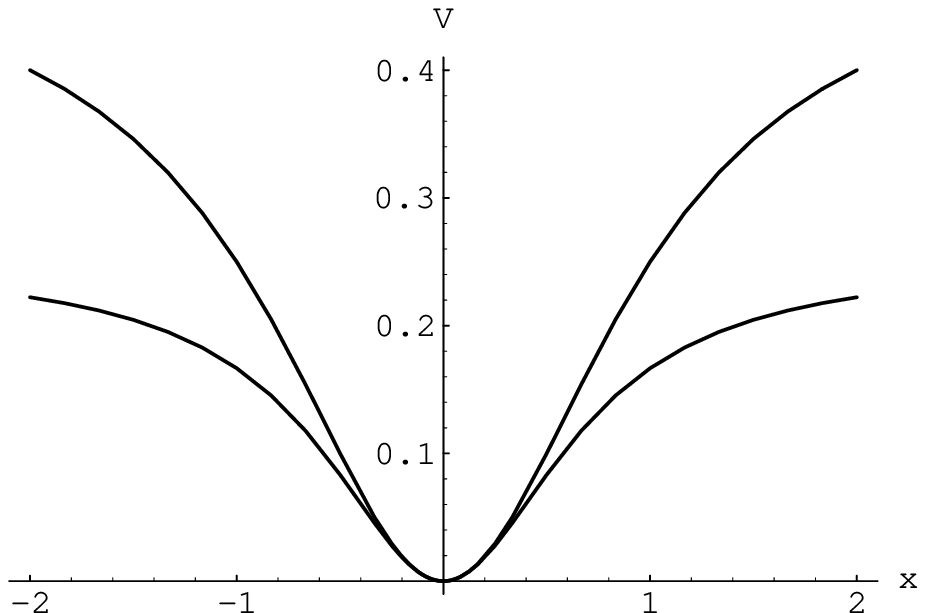}
$$

{\sc Figure II}.{\enskip}
Plot of $V_\la(x)=(1/2)\,(\al^2 x^2)/(1 + \la\,x^2)$, $\al=1$, $\la>0$,
as a function of $x$, for $\la=1$
(upper curve), and $\la=2$ (lower curve).

\vfill\eject

$$
   \epsfbox{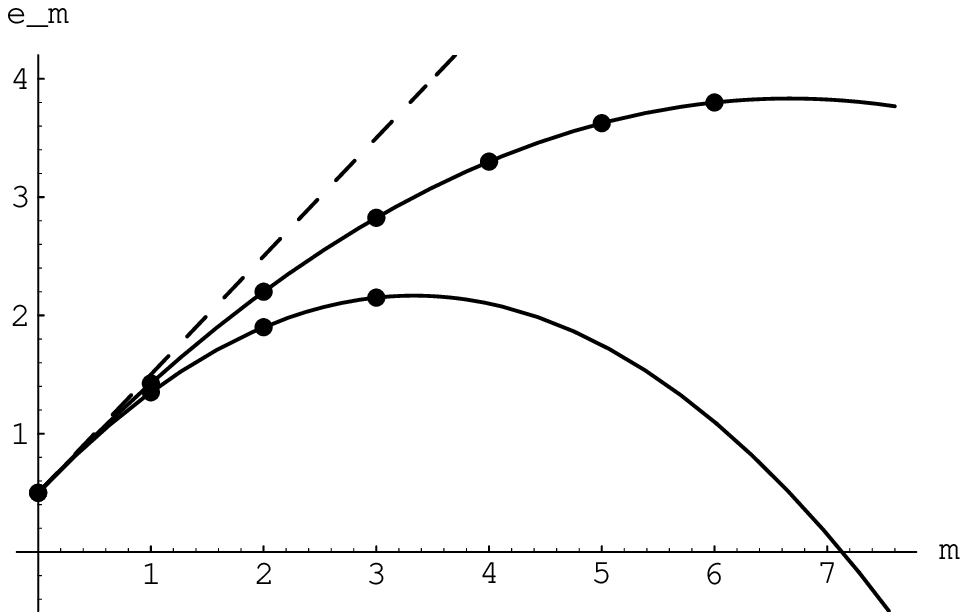}
$$

{\sc Figure III}.{\enskip} Plot of the energy $e_m$ as a function
of $m$ for $\La=0.30$ (lower curve) and $\La=0.15$ (upper curve).
The curves also show the plot of the points $(m,e_m)$ for the
values $m=0, 1, 2, 3$, and $m=0, 1, \dots, 6$, respectively. For
$\La=0.30$ there exist four bound states (four thick points in the
curve) and for $\La=0.15$ seven bound states (seven thick points
in the curve). When $\La$ decreases the maximum of the curve moves
into the up right and in the limit $\La\to 0$ the curve converges
into a straight line parallel to the diagonal (dashed line).

{\vskip 40pt}
$$
    \epsfbox{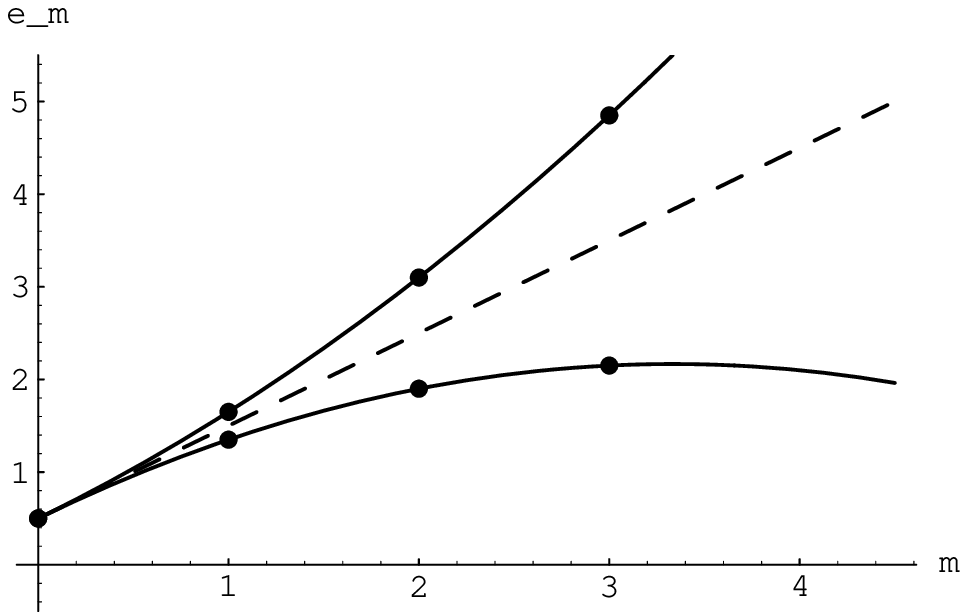}
$$

{\sc Figure IV}.{\enskip}
Plot of the energy $e_m$ as a function of $m$ for $\La=0.30$
(lower curve) and  $\La=-0.30$ (upper curve).
The thick points $(m,e_m)$, corresponding to the values
$m=0, 1, 2, 3$, represent the four bound states existing
for $\La=0.30$ and the first four bound states for $\La=-0.30$.
The straight line (dashed line) placed in the middle corresponds
to the linear harmonic
oscillator.

\end{document}